\documentclass[12pt]{article}
\usepackage{amsmath,amsfonts,amssymb,latexsym}
\usepackage{epsfig}
\usepackage{graphicx}
\usepackage{wrapfig}
\usepackage{subfigure}
\usepackage{cite}

\def\be{\begin{equation}}
\def\ee{\end{equation}}
\def\bea{\begin{eqnarray}}
\def\eea{\end{eqnarray}}
\def\bean{\begin{eqnarray*}}
\def\eean{\end{eqnarray*}}
\def\N{{\mathbb N}}

\def\R{{\mathbb R}}
\def\C{{\mathbb C}}
\def\I{{\mathbb I}}

\begin{document}


\thispagestyle{empty}
\hfill \today

\vspace{.8cm}

\begin{center}
\bf{\LARGE 
Coherent Orthogonal Polynomials}
\end{center}

\bigskip

\begin{center}
E. Celeghini$^1$  and M.A. del Olmo$^2$
\end{center}

\begin{center}
$^1${\sl Departimento di Fisica, Universit\`a  di Firenze and
INFN--Sezione di
Firenze \\
I50019 Sesto Fiorentino,  Firenze, Italy}\\
\medskip
$^2${\sl Departamento de F\'{\i}sica Te\'orica, Universidad de
Valladolid, \\
E-47005, Valladolid, Spain.}\\
\medskip
{e-mail: celeghini@fi.infn.it, olmo@fta.uva.es}

\end{center}

\bigskip

\begin{abstract}

 We discuss as a fundamental characteristic of
orthogonal polynomials like the existence of a  Lie algebra behind them, can be 
added to their other relevant aspects.
At the basis of the complete framework for orthogonal polynomials we put thus --in addition to
differential equations, recurrence relations, 
Hilbert spaces and square integrable functions-- Lie algebra theory. 

We start here from the square integrable functions on the open connected subset of the real line 
whose bases are related to orthogonal polynomials.
All these one-dimensional continuous spaces allow, 
besides the standard uncountable basis $\{|x\rangle\}$, for an 
alternative countable basis $\{|n\rangle\}$.
The matrix elements that relate these two bases are essentially the orthogonal polynomials:
Hermite polynomials for the line and  Laguerre and Legendre polynomials for the half-line and the line  interval, respectively.

Differential recurrence relations of orthogonal polynomials allow us to realize that they determine
a unitary representation of a non-compact Lie algebra, whose  second order Casimir ${\cal C}$ gives rise to the 
second order differential
equation that defines  the corresponding family of orthogonal polynomials. Thus, the Weyl-Heisenberg algebra  
$h(1)$ with ${\cal C}=0$ for Hermite polynomials and $su(1,1)$ with 
${\cal C}=-1/4$\,  for 
Laguerre and Legendre polynomials are obtained.

Starting from the orthogonal polynomials the Lie algebra
is extended both to the whole space of the ${\cal L}^2$ functions and to
the corresponding Universal Enveloping Algebra and transformation group.
Generalized coherent states from each vector in the space ${\cal L}^2$ and, in particular, generalized
coherent polynomials are thus obtained. 
\end{abstract}
\vskip 1cm 

Keywords: orthogonal polynomials, group representation theory
\smallskip

PACS: 02-20.Qs, 02-20.Sv, 02-30.Gp, 03-65.Fd,  

\vfill\eject

\section{Introduction}

Orthogonal polynomials are relevant in many fields of mathematics: differential equations,
 algebras and Lie groups, Hilbert spaces and generalized Fourier series are, perhaps, the principal ones.

They are essentially orthogonal bases of square integrable functions ${\cal L}^2(E)$.  
We consider in this paper the cases discussed by Cambanis in
 \cite{Camb71}, 
where $E$ is a connected open subset of the real line,  i.e. $E=(a,b)\subset \R$ 
with $-\infty \le  a < b \le + \infty$. We will deal with  the Hermite
functions when $E=(-\infty,+\infty)$, the Laguerre functions for $E = (a,+\infty)$ with $-\infty <a$ or 
$E = (-\infty, b)$ with $b<+\infty$  (both with an appropriate change of variables reduce
to $E = (0,+\infty)$) and the Legendre polynomials for $E = (a,b)$  with $-\infty <  a < b < + \infty$  
(that, with a suitable change of variables, reduces to $E = (-1,+1)$).
We will see that the operators acting on each vector space  have a Lie
algebra structure.

The framework we are dealing with is related to the intertwining of the basis that defines 
the space and the alternative countable basis introduced by the orthogonal polynomials.
The generators of the Lie algebra are shown to be related to
first order differential recurrence relations 
while ${\cal C}$, the second order invariant of the algebra, originates the second order 
differential equation
that defines the family of orthogonal polynomials.
From the generators, the full set of operators on the space ${\cal L}^2(E)$ is built inside the Universal Enveloping Algebra 
(UEA), i.e. the closure of the underlying vector space  of a Lie algebra that has  the ordered monomials constructed 
on the generators as a basis.
In particular, as the group elements belong to the UEA,
the group transformations on ${\cal L}^2(E)$ are in deep relationship with the orthogonal polynomials.
Specifically, each type of orthogonal polynomials is related to a unitary irreducible  representation (UIR) of a
non-compact Lie group: Hermite functions (as it is well-known from the quantum harmonic oscillator) support a UIR
of the Weyl-Heisenberg group $H(1)$ and the Laguerre
and Legendre functions belong to the fundamental representation of the discrete series of $SU(1,1)$.

As algebra and group operators are defined on the 
whole ${\cal L}^2(E)$ space  we can obtain
in an operatorial way peculiar combinations of vectors of 
${\cal L}^2(E)$\, as, for instance, 
generalized coherent orthogonal polynomials.    

Our starting point is the separable Hilbert space ${\cal L}^2(E)$ equipped with the basis\, $\{|x\rangle\}$\,
on the connected open subset $E=(a,b)\subset \R$. 
Orthonormality and completeness are
\be\label{ortcomx}
\langle \,x\,|\,y\,\rangle\, =\, \delta(x-y), \qquad
 \int_{a}^{b}\,  |x\rangle \,dx\, \langle x|\, =\, \mathbb I.
\ee
For each type of orthogonal polynomials we will write first the vector space --essentially to
fix the notations-- and then we will discuss the operators acting on it.

\section{Hermite polynomials}
 
We start our discussion from the Hilbert space constructed by 
means of the Hermite polynomials $H_n(x)$, defined on the full real line $(a,b) = (-\infty, \infty)$,
because in this case both, vector space and  Lie algebra 
structure, are
well known and we shall limit ourselves to put together, in a compact way, known results. 
 
To begin with, some factors are attached to the $H_n(x)$, writing the usually called 
Hermite functions $K_n(x)$:
\be\label{Kfunctions}
K_n(x)\,:=\; \frac{e^{-x^2/2}}{\sqrt{2^n n! \sqrt{\pi}}}\, H_n(x)\, ,
\qquad n\in \N .
\ee
In terms of them orthonormality and completeness are\cite{thooft}
\be \begin{array}{l}\label{Kfunctionsrelations}\displaystyle
 \int_{-\infty}^{\infty}\,  K_{n}(x)\; K_{m}(x) \;dx\;=\;\delta_{n,m}\;, \\[0.4cm]
\displaystyle \sum_{n=0}^{\infty}\,   K_{n}(x)\; K_{n}(y)\; =\; \delta(x-y)\;.
\end{array}\ee
Hence, the functions $\{K_n(x)\}$ determine an orthonormal basis for the real square integrable 
functions on the line, ${\cal L}^2((-\infty,\infty))={\cal L}^2(\R)$ \cite{Szek}.
We can now define the set $\{|n\rangle\}_{n\in \N}$: 
\be\label{n}
 |n\rangle\,:=\;\; \int_{-\infty}^{\infty}  |x\rangle\; K_n(x)\; dx\;, 
\ee
that is 
isomorphic to $\{K_n(x)\}$ and also an orthonormal basis since from  
\eqref{ortcomx}, \eqref{Kfunctions} and \eqref{Kfunctionsrelations}
 \be\nonumber
\langle n| m\rangle = \delta_{n,\,m}\,,\qquad
 \sum_{n=0}^{\infty}\, |n\rangle\, \langle n|\, =\, \I \label{ortcomn} \,.
\ee
Note that, instead of the usual heavy notation\, 
$|K_n(x)\rangle$\,,
we use the physical notation $|n\rangle$ to emphasize the two bases 
$\{|x\rangle\}_{x\in\R}$ and $\{|n\rangle\}_{n\in\N}$ of the   Hilbert space ${\cal L}^2(\R)$.
Orthogonal functions play the role of transition matrices
and, as the space is real, are written as
\be\nonumber
K_n(x) = \langle x| n \rangle = \langle n| x\rangle .
\ee
Now, in complete analogy with the 
Fourier analysis on the circle, an arbitrary vector $|f\rangle$ of 
 ${\cal L}^2(\R)$
can be written
\be\label{f}
|f\rangle = \int_{-\infty}^{+\infty}\,  |x\rangle \,dx\, \langle x|f\rangle \quad = \quad 
\sum_{n=0}^\infty |n\rangle \langle n| f \rangle \, ,
\ee
where the wave function $f(x)$ and the sequence $\{f_n\}_{n\in\N}$
describe the vector $|f\rangle$ in the two bases:
\be\label{fxn}
f(x):= \langle x|f\rangle = \sum_{n=0}^\infty K_n(x) f_n \qquad 
f_n:= \langle n  | f\rangle = \int_{-\infty}^{+\infty} K_n(x) f(x) dx .
\ee
In particular, the completeness of the two bases determines the inner
product as well as the Parseval identity:
\be\label{Parseval}
\langle g | f \rangle = \sum_{n=0}^\infty g_n\cdot f_n \; = 
\int_{-\infty}^{+\infty}  g(x) \cdot f(x) \; dx \,,\qquad
\sum_{n=0}^\infty f_n^{\,2} = \int_{-\infty}^{+\infty} f(x)^2\, dx\,.
\ee

Let us introduce now the operators defined on the line that,  as
it is well-known, are functions of creation, annihilation and number operators. 
Following the factorization method\cite{infeld-hull1951,miller1968} we consider, among
the recurrence relations of Hermite polynomials, those that include first
order derivatives \cite{thooft}:
\be\label{HP}
H'_n(x) = 2\, n\, H_{n-1}(x), \qquad H'_n(x) -2\, x\, H_n(x) = - H_{n+1}(x)\;.
\ee 
The second order differential equation, that defines the Hermite polynomials,
is obtained by subsequent applications of recurrence relations.
However, the fundamental limitation of such an approach is  
that in the factorization method the problem has been considered  from
the point of view of differential equations theory and, hence,  
the labels are assumed to be parameters. 

We shall show here that a consistent Hilbert space 
framework, where the operators are well defined, allows us 
to reformulate special functions in the Lie algebra representation scheme, such that, in particular,
Hermite functions 
support a well defined UIR of 
the Weyl-Heisenberg algebra $h(1)$.
Effectively, we can rewrite eqs.(\ref{HP}) in terms of Hermite functions obtaining
\be\label{reckn}
a\; K_n(x)\; =\; \sqrt{n}\, K_{n-1}(x) \,,\qquad 
a^\dagger\, K_n(x)\; =\; \sqrt{n+1}\,\; K_{n+1}(x)\, ,
\ee
where $a$ and $a^\dagger$ are the annihilation and creation operators of the
harmonic oscillator:
\be\label{aa+}
a:= \frac{1}{\sqrt{2}} \left (X + D_x\right )\;,\qquad 
a^\dagger := \frac{1}{\sqrt{2}} \left ( X - D_x \right )\, .
\ee
In addition to the two previous
operators $X$ and $D_x$ 
\be\label{xdx}
X\, f(x) := x\, f(x)\;, \qquad D_x\, f(x) := f'(x)\,, 
\ee
we introduce  two other operators $N$ and $I$ defined by
\be\label{nnn}
N\, K_n(x) := n\, K_n(x)\,,\qquad I\, K_n(x) :=  K_n(x)\, ,
\ee
in order to complete our set of operators.
Note that the change  introduced by substituting  the label $n$ 
of the Hermite polynomials for the operator $N$ 
is far to be irrelevant since $N$ does not commute with  $a$ and $a^\dagger$:
\[
[N,a]= - a \,,\qquad [N,a^\dagger ] = a^\dagger\, .
\]
The whole Weyl-Heisenberg algebra follows
\[
[N,a]=-a ,\qquad [N,a^\dagger] = a^\dagger , \qquad\qquad 
[ a, a^\dagger ] = I ,
\qquad\qquad  [ I, \bullet \,] = 0\;.
\]
Expression (\ref{n}) allows us to connect to
the operators $a$, $a^\dagger$, $N$ and $I$, defined on the set $\{K_n(x)\}$,
four related operators defined on the line Hilbert space that we denote 
with the same symbols:
\[
a\, |n\rangle \,=\, \sqrt{n}\;\, |n-1\rangle, \qquad 
a^\dagger\, |n\rangle  \,=\, \sqrt{n+1}\;\, |n+1\rangle ,\qquad 
N\, |n\rangle = n |n\rangle , \qquad 
 I\, |n\rangle = |n\rangle .
\]
 The quadratic invariant operator ${\cal C}$ can be written \cite{ceta}
\be\nonumber
{\cal C}\; = \; \{a, a^\dagger \} - 2 (N+1/2)\, I .
\ee
As, by inspection, we are dealing with the representation 
${\cal C} = 0$
the Schr\"odinger equation 
of the quantum harmonic oscillator is obtained, by means of eqs.(\ref{reckn}--\ref{nnn}), i.e.  
\be\nonumber
{\cal C} \; K_n(x)\equiv \left( X^2 - D_x^2 - (2 N+1)\right )\; K_n(x) = 0 ,
\ee
that, rewritten in terms of the Hermite polynomials allows us to recover the equation defining  the Hermite polynomials
\be\nonumber
H''_n(x) - 2\,x\, H'_n(x) + 2\,n\, H_n(x) = 0\, .
\ee
Thus, we can conclude that the Hermite functions constitute a basis of the UIR ${\cal C}=0$ of
the Weyl-Heisenberg algebra\, $h(1)$ and the whole UEA of $h(1)$ is  defined
on the $\{K_n(x)\}$. In particular, as all elements of the Weyl-Heisenberg group $H(1)$ are
contained inside the UEA of $h(1)$, all the transformations of $H(1)$ can be
realized in the space $\{K_n(x)\}$. As an example, eqs.(\ref{reckn}) allow to obtain the simplest normalized
coherent Hermite function as\cite{perelomov}:
\be\nonumber
K_{z}(x) :=\,\frac{1}{\sqrt{e^{|z|^2}}} \; e^{\alpha\, a^\dagger}\, K_0(x)=
\frac{1}{\sqrt{e^{|z|^2}}}\sum_{n=0}^{\infty}\frac{z ^n}{\sqrt{n!}}\,K_n(x), \qquad z\in \C\,.
\ee 

\section{Laguerre polynomials}

The Laguerre functions differ from Hermite ones in the relevant fact
that recurrence relations depend  not only on the variable $x$ (as it happens for $a$ 
and $a^\dagger$ in eqs.(\ref{aa+})) but also on the parameter $n$.

The discussion is  similar to the previous one.
We shall start renormalizing the Laguerre polynomials $L_n(x)$ by
defining in this way the Laguerre functions
\be\nonumber
M_n(x) := e^{-x/2}\,  L_n(x) .
\ee
Orthonormality and completeness relations are \cite{Camb71}:
\be \begin{array}{l}\label{Mfunctionsrelations}\displaystyle
 \int_{0}^{\infty} M_{n}^{}(x)\, \, M_{m}^{}(x) \;dx\;=\; \delta_{n,m}\,,\\[0.4cm]
\displaystyle \sum_{n=0}^{\infty}  \;M_{n}^{}(x)\,\, M_{n}^{}(y)\; =\; 
\delta(x-y). 
 \end{array}\ee
In analogy with expression (\ref{n}),
expressions (\ref{ortcomx}) allows us to define
\be\nonumber
 |n\rangle\,:=\;\; \int_{0}^{\infty}  |x\rangle\; M_n(x)\; dx \,,
\ee
and to obtain a countable basis $\{|n\rangle\}_{n\in \N}$ of the half-line, 
in addition to the standard basis $\{|x\rangle\}_{x\in (0,\infty)}$. 
The  functions $M_n(x)$ are the transformation matrix elements that relate the two orthonormal bases 
\be\nonumber
M_{n}^{}(x)\,=\; \langle \,x\,|\,n\,\rangle\; =\;  \langle \,n\, |\,x\,\rangle\,,
\ee
and for an arbitrary vector $|f\rangle$ of the Hilbert 
space ${\cal L}^2((0,\infty))$ 
all formulas (\ref{f}-\ref{Parseval}) are valid  after the changes  $(-\infty,+\infty)\to (0,+\infty)$ and $\{K_n(x)\} \to 
\{M_n(x)\}$.

In order to obtain the operators acting on ${\cal L}^2((0,\infty))$, let us write the recurrence formulas 
involving derivatives for the Laguerre polynomials \cite{NIST}
\be\nonumber
x\, L'_n(x) -n\, L_n(x) = -n\, L_{n-1}(x)\,, \qquad x\, L'_n(x) +(1+n-x)\, L_n(x) = (n+1)\, L_{n+1}(x)\,,
\ee
that can be rewritten on terms of the functions $M_n(x)$ in a quite more symmetrical form
\be \begin{array}{l}\label{Moperators}\displaystyle
J_-\; M_n(x):= \left(-x \frac{d}{dx} +n -\frac{x}{2}\right) M_n(x) = n\; M_{n-1}(x)\,, 
\\[0.4cm] \displaystyle
J_+\; M_n(x) := \left(x \frac{d}{dx} +n+1 -\frac{x}{2}\right) M_n(x) = (n+1)\; M_{n+1}(x)\, .
\end{array}\ee
Because the operator $X D_x$ is zero for $x=0$, we 
can properly define $J_\pm$ as true operators on the half-line Hilbert space 
\be\label{J-+RR}
J_-\, := -X D_x +N -\frac{X}{2} \,,\qquad
J_+\, := X D_x +N+1 -\frac{X}{2}\,.
\ee 
Like in the Hermite case, the introduction of these full operatorial structures
is not trivial since the operators 
$N$ and $J_\pm$ do not commute  
\be\nonumber
[N,J_\pm]=\pm J_\pm\, .
\ee
The operators acting on the basis $\{|n\rangle\}$ are equivalent to the ones acting on the basis $\{M_n(x)\}$
\be\label{j-+lr}
J_-\, |n\rangle = n\; |n-1\rangle\,,\qquad J_+\, |n\rangle = (n+1)\; |n+1\rangle\,,\qquad 
N\, |n\rangle = n\; |n\rangle\,.
\ee
Defining\, $J_3:= N+1/2$\,,
commutators and anticommutators of $J_\pm$ are
\be\nonumber
[\,J_+,\,J_-]\, =\, -2\, J_3\,, \qquad \{\,J_+,\, J_-\} = 2\, J_3^2\,+1/2\,.
\ee
and, 
$\{M_n(x)\}$ and $\{|n\rangle\}$ are 
found to be representations of the $su(1,1)$ algebra:
\be\nonumber
[J_3, J_\pm] = \pm J_\pm \qquad [J_+, J_-] = -2 J_3\,.
\ee

The differential equation defining the Laguerre polynomials can be obtained, following the 
factorization metod, from
\[(J_+ J_- - N^2)\, M_n(x)=0\,,
\] 
or 
\[ 
(J_- J_+ - (N+1)^2)\, M_n(x) = 0\,.
\]
Both expressions, using eqs. (\ref{J-+RR}), give 
\be\nonumber
- X \left(X D^2_x + D_x +N +1/2 -X/4\right)\, M_n(x) =0\,,
\ee
which is equivalent, since $X$ in never the null operator on $(0,+\infty)$, to
\be\label{C2Lrr}
\left(X D^2_x + D_x +N +1/2 -X/4\right) M_n(x) =0\, ,
\ee
that, rewritten in terms of the $\{L_n(x)\}$,
is the definition equation of the Laguerre polynomials, i.e. 
\be\nonumber
x\, L''_n(x)+ (1-x)\, L'_n(x) +n\, L_n(x) = 0\, .
\ee

Alternatively, following the algebraic approach, the anticommutator $\{\,J_+,\, J_-\}$
is related to the Casimir operator ${\cal C}$. From eqs.(\ref{j-+lr})
\be\nonumber
{\cal C}= \left(J_3^2-\frac 12 \{J_+,J_-\}\right)\, =\, - \frac14\;\, ,
\ee
that, introducing expressions (\ref{J-+RR}), gives the operational form of eq.(\ref{C2Lrr}), i.e. the 
definition of Laguerre polynomials.

Since\, ${\cal C} = -1/4$\, and\, $J_\pm^\dagger = J_\mp$\,
we are dealing with the fundamental representation of the discrete series of UIR of $su(1,1)$. 
Because\, ${\cal C}=k \;(k-1)$ with  $k$ the maximum weight of the representation, we find $k=1/2$
in agreement with the minimum eigenvalue $0$ of the operator $N$ of the Laguerre
polynomials.  Hence, the spectrum of the operator $J_3$  is $1/2,\, 3/2,\, 5/2,\dots$

Again,  from the algebra $su(1,1)$ we can move to its UEA and to the group $SU(1,1)$. In this way, 
a coherent Laguerre polynomial, $L_{\alpha}(x)$, could thus be defined  \`a la 
Perelomov \cite{perelomov}.  These coherent states are in correspondence with the points of the upper sheet of the two-sheet hyperboloid $\mathbb H=\{ (y_0,y_1,y_2) \,|\, y_0^2-y_1^2-y_2^2=1,\; y_0>0\}$, and  we can parametrize the points of  $\mathbb H$ by the coordinates $(\xi,\theta)$, with $ \xi \in \R^+$ and $\theta \in [0,2 \pi)$, as follows: 
$y_0=\cosh \xi,\;y_1=\sinh \xi \cos \theta,\;y_2=\sinh \xi \sin \theta$. Then 
\be\begin{array}{lcl}\label{cslaguerre}
L_{\alpha}(x)& :=&\; e^{x/2}\; e^{\hat \xi J_+ - \hat \xi^* J_-}\; M_0(x) =\; e^{x/2}\;
 e^{\alpha J_+}\; (1-|\alpha |^2)^{J_3}\;e^{ - \alpha^* J_-}\; M_0(x)\\[0.35cm] 
&=& e^{x/2}\; (1-|\alpha |^2)^{1/2}\;e^{\alpha J_+}\; M_0(x)\\[0.35cm]
&=& \displaystyle
 (1-|\alpha |^2)^{1/2}\;\sum_{n=0}^{\infty} \alpha^n\; L_n(x)\;,
 \end{array}\ee
where $ \hat \xi=\xi e^{i \theta}$ and $ \alpha= e^{i \theta}\tanh \xi $.

\section{Legendre polynomials}

The approach is similar to that of Laguerre polynomials, with the difference that the interval $E=(a,b)\subset \R$ is now $(-1,1)$. 
We have \cite{thooft}:
\be \begin{array}{l}\label{ortLP}\displaystyle
 \int_{-1}^{1} P_{n}^{}(x)\, (n+1/2)\, P_{m}^{}(x) \;dx\;=\; \delta_{n,m}\, ,
 \\[0.4cm] \displaystyle
 \sum_{n=0}^{\infty}  \;P_{n}^{}(x)\, (n+1/2)\, P_{n}^{}(y)\; =\; \delta(x-y).
\end{array}\ee
As before expressions (\ref{ortLP}) allow us to define a new basis $\{|n\rangle\}_{n\in \N}$ 
in the interval $(-1,1)$ in addition to the standard basis $\{|x\rangle\}_{x\in (-1,1)}$ :
\be\nonumber
 |n\rangle\,:=\;\; \int_{-1}^{+1}  |x\rangle\;\sqrt{n+1/2} \; P_n(x)\; dx \,.
\ee 
The  $P_n(x)$ polynomials are thus the transformation matrix elements that relate both two orthonormal bases: 
\be\nonumber
\sqrt{n+1/2}\; P_{n}^{}(x)\,=\; \langle \,x\,|\,n\,\rangle\; =\;  \langle \,n\, |\,x\,\rangle \,,
\ee
and, in complete analogy with the previous cases, for an arbitrary vector 
$|f\rangle$ of the Hilbert 
space ${\cal L}^2((-1,+1))$ 
all formulas (\ref{f}-\ref{Parseval}) are valid for the Legendre case after the changes  $(-\infty,+\infty)\to (-1,+1)$ and $\{K_n(x)\} \to  \{\sqrt{n+1/2}\;P_n(x)\}$.

The vector space structure described, let us now study to the operators. The recurrence
formulae \cite{NIST}:
\be \begin{array}{l}\label{recurrenceLP}\displaystyle
 (x^2-1)\; P_n^{'}(x)= ( n+1)\;  \left( P_{n+1}^{}(x)-x\;P_n^{}(x) \right)\, , 
\\[0.3cm] \displaystyle
 (x^2-1)\; P_n^{' }(x)=  n\; ( x\; P_n^{}(x)- P_{n-1}^{}(x))\,,
 \end{array}\ee
can be rewritten in terms of  
differential operators as follows:
\be\label{j-+Leg}\begin{array}{l} 
J_- \, P_n^{}(x) \equiv  \displaystyle\left( (1-X^2) D_x +X\; N\right)\, P_n^{}(x) 
= n\, P_{n-1}^{}(x)\,,\\[0.3cm]
J_+ \,P_n^{}(x) \equiv  \displaystyle\left( -(1-X^2) D_x +X \,(N+1)\right)\, 
P_n^{}(x)  = (n+1)\, P_{n+1}(x)\, . 
\end{array}\ee
Again, the change we introduced into 
eqs. (\ref{j-+Leg}),  is not irrelevant since
\[
[N,J_\pm]=\pm J_\pm  , 
\]
and, also in this case,the hermiticity relation\, $(J_\pm)^\dagger=J_\mp $ (and, thus, the unitarity of the representation)
is imposed by the recurrence formulas. 

The action of the operators $J_\mp $ on the vectors $| n\rangle$ is from \eqref{j-+Leg}
\be\label{Jactions}
  J_-\; | n\rangle= n\; |n-1\rangle\,,  \qquad
 J_+\;  |n\rangle= (n+1) \; |n+1\rangle\ .
\ee
Finally from   from \eqref{Jactions} 
 we obtain:
\be\nonumber\begin{array}{l}
J_+J_- -N^2=- (1-X^2)\;\left( (1-X^2) D_x^2-2 X D_x+N(N+1)\right)\equiv 0,\\[0.3cm]
J_-J_+ -(N+1)^2=- (1-X^2)\;\left( (1-X^2) D_x^2-2 X D_x+N(N+1)\right)\;\equiv 0 ,
\end{array}\ee
that, up to an irrelevant global factor $(1-X^2)$ that  never vanishes,
is the Legendre equation in operatorial form.

As before, the algebraic approach starts from the commutator and the anticommutator of the operators $J_\pm$
\be\nonumber
[J_+,J_-]\; = -\, 2\left(N+1/2\right)\;\, ,\qquad
\{J_+,J_-\}\; = 2  (N+1/2)^2+1/2 \, .
\ee
Defining the operator $J_3:\,=\; N+1/2$, which has the same explicit form than in   
the Laguerre case,  a 
$su(1,1)$ algebra is again found:  
\be\nonumber
[J_3,J_\pm]= \pm  J_\pm, \qquad
[J_+,J_-]=- 2 J_3\,. 
\ee
Moreover, the Legendre functions support the same UIR of Casimir ${\cal C} = -1/4$ of  the Laguerre functions
and the equation
\[
({\cal C} + 1/4)\; P_n(x)\,= \left(J_3^2- \{J_+,J_-\}/2\, + 1/4\right)\;P_n(x)\,= \;0
\]
togheter with eqs.(\ref{j-+Leg}) gives the   Legendre equation.

A coherent state generalization of $\{P_n(x)\}$ can be also defined in this case similarly to \eqref{cslaguerre} 
\be\label{cslegendre}
P_{\alpha}(x) := e^{\hat \xi J_+ - \hat \xi^* J_-}\; P_0(x) 
= \displaystyle
 (1-|\alpha |^2)^{1/2}\;\sum_{n=0}^{\infty} \alpha^n\; P_n(x)\;,
\ee 
\noindent where the complex numbers $\alpha$ and  $\hat \xi$ are the same of the Laguerre case 
\eqref{cslaguerre}.

\section{Conclusions}

This paper is the first of a program attempting to give a global description of orthogonal polynomials
where the different aspects --differential equations theory,
Lie algebras, Hilbert spaces and generalized Fourier series--  are considered together. 
We discuss here orthogonal polynomials defined on the line --where technical
aspects are less relevant-- and we give an algebraic description of them in a global way.
At the level of Hilbert spaces the Hermite polynomials allow to build a discrete basis
from the square integrable functions on the line while the Laguerre and Legendre polynomials are defined
on the half-line and the interval $(-1,+1)$, respectively.
From the point of view of Lie algebras, the Hermite polynomials support the UIR of the Weyl-Heisenberg  algebra associated to the Casimir
${\cal C} = 0$ and the Laguerre and Legendre
polynomials are related to the fundamental UIR (${\cal C} = -1/4$) of the discrete series of $su(1,1)$ that are all explicitly constructed.

Summarizing, in all the three cases we introduce 
in the vector space the countable basis 
$\{|n\rangle\}_{n\in \N}$, related to the uncountable coordinate basis
$\{|x\rangle \}_{x\in E\subset \R}$ by the appropriate orthogonal functions, on which a UIR
 of a non-compact Lie algebra (the Weyl-Heisenberg algebra $h(1)$ for
Hermite and the $su(1,1)$ algebra for Laguerre and Legendre) is defined. As 
the basis $\{|n\rangle\}$
and the associated orthogonal functions are equivalent the algebraic structure
can be transferred from the ${\cal L}^2(E)$ space on the vector space straight to the vector space $E$. 
Moving from the Lie algebra to the UEA, all the operators defined on the  space on 
${\cal L}^2(E)$ can be generated. In
view of applications it
is particularly relevant the action on ${\cal L}^2$ of the groups $H(1)$ or $SU(1,1)$.    
A first result related to the introduction of the algebraic structure is that
coherent states defined on the the line vector space $\{|n\rangle\}$ by the Weyl-Heisenberg 
algebra can be transferred to the wave functions, i.e. the Hermite polynomials and, analogously,
coherent Laguerre or Legendre polynomials can be defined \`a la 
Perelomov \cite{perelomov} 
starting from the properties of the appropriate vector space. 

In conclusion, we belive that --combining properties of different origin like differential equations, 
factorization method, theory of Lie representations, Hilbert spaces, integrable functions--
a better description of the orthogonal polynomials has been obtained.

\section*{Acknowledgments}

This work was partially supported  by the Ministerio de
Educaci\'on y Ciencia  of Spain (Projects FIS2009-09002 with EU-FEDER support),  by the 
Junta de Castilla y Le\'on and by
INFN-MICINN (Italy-Spain).



\begin{thebibliography}{99}

\bibitem{Camb71} S. Cambanis, {\it Proc. of the Am. Math. Soc.} {\bf 29},
284 (1971).

\bibitem{thooft}
G. 'tHooft and S. Noobbenhuis,\, {\it Special Functions and Polynomials}\hfill\vspace{-.55cm}
\begin{verbatim}
www.phys.uu.nl/~hooft101/GtH_lectures.html.
\end{verbatim}

\bibitem{Szek} P. Szekeres, {\it A Course in Modern Mathematical Physics}
(Cambridge: Cambridge Univ. Press, 2006).

\bibitem{infeld-hull1951}  
L. Infeld and T.E. Hull, {\it Rev. Mod. Phys}. 
{\bf 23}, 21 (1951). 

\bibitem{miller1968}
W. Miller, {\it Lie Theory and Special Functions} (New York: Academic Press, 1968).

\bibitem{ceta}
E. Celeghini and M. Tarlini, {\it Il Nuovo Cimento} {\bf B61}, 265 (1981).

\bibitem{perelomov} A. Perelomov, {\it Generalized Coherent States and Applications}
(Berlin: Springer, 1986).

\bibitem{NIST}
F.W.J. Olver, D.W. Lozier, R.F. Boisvert and C.W. Clark, 
{\it NIST Handbook of Mathematical Functions} (New York: Cambridge Univ. Press, 2010). 
 
\end{thebibliography}
\end{document}